\documentclass[apjl,letter]{emulateapj}

\usepackage{times}

\begin{document}

%% \title{Spiral arms in transition disks as a consequence of cast shadows}
\title{Spiral waves triggered by shadows in transition disks}

\author{Mat\'ias Montesinos$^{1,2}$}
\author{Sebastian Perez$^{1,2}$}
\author{Simon Casassus$^{1,2}$}
\author{Sebastian Marino$^{1,2}$}
\author{Jorge Cuadra$^{3,2}$}
\author{Valentin Christiaens$^{1,2}$}

\affil{$^{1}$Departamento de Astronom\'ia, Universidad de Chile,
  Casilla 36-D, Santiago, Chile; montesinos@das.uchile.cl}
\affil{$^{2}$ Millennium Nucleus ``Protoplanetary Disks'', Chile}
\affil{$^{3}$ Instituto de Astrof\'isica, Pontificia Universidad
  Cat\'olica de Chile, Santiago, Chile}

%\date{\today}
%\maketitle

%label{firstpage}

\begin{abstract}

Circumstellar asymmetries such as central warps have recently been
shown to cast shadows on outer disks. We investigate the
hydrodynamical consequences of such variable illumination on the outer
regions of a transition disk, and the development of spiral
arms. Using 2D simulations, we follow the evolution of a gaseous disk
passively heated by the central star, under the periodic forcing of
shadows with an opening angle of $\sim$28$^\circ$. With a lower
pressure under the shadows, each crossing results in a variable
azimuthal acceleration, which in time develops into spiral density
waves. Their pitch angles evolves from $\Pi \sim 15^\circ-22^\circ$ at
onset, to $\sim$11$^\circ$-14$^\circ$, over $\sim$65~AU to
150~AU. Self-gravity enhances the density contrast of the spiral
waves, as also reported previously for spirals launched by planets.
Our control simulations with unshadowed irradiation do not develop
structures, except a different form of spiral waves seen at later
times only in the gravitationally unstable control case.  Scattered
light predictions in $H$ band show that such illumination spirals
should be observable.  We suggest that spiral arms in the case-study
transition disk HD~142527 could be explained as a result of shadowing
from the tilted inner disk.
\end{abstract}


%\onecolumn

\section{Introduction}

During the last decade optical-infrared direct imaging of
circumstellar disks has revealed spiral patterns around some HAeBe
stars. These spirals are seen in intermediate mass stars,
$\sim$2~M$_\odot$, with an inner cavity in a gas-rich disk, and are
loosely classified as transition disks \citep[e.g.][]{Espaillat2012}.
Outstanding examples are AB~Aur
\citep[e.g.][]{Grady1999ApJ...523L.151G, Fukagawa2004ApJ...605L..53F},
HD~100546 \citep[e.g.][]{Grady2001AJ....122.3396G,
  Boccaletti2013AA...560A..20B}, HD~142527
\citep[e.g.][]{Fukagawa2006, Casassus2012ApJ...754L..31C}, MWC~758
\citep[e.g.][]{Grady2013ApJ...762...48G, Benisty2015AA...578L...6B},
HD~135344B \citep[e.g.][]{Muto2012ApJ...748L..22M,
  Garufi2013AA...560A.105G, Wahhaj2015AA...581A..24W}, and HD~100453
\citep[][]{Wagner2015ApJ...813L...2W}. Spirals usually stem away from
the outer rims of disk cavities, with large pitch angles
\citep[$10^\circ-15^\circ$,][]{Dong-2015a}. They can extend from
$\sim$15~AU to 600~AU from the central star
\citep{Clampin2003AJ....126..385C, Christiaens-2014} and some of them
show remarkable $m=2$ azimuthal symmetry \citep{Dong-2015a}.

The origin of such spirals is motivating intense research efforts.
Spiral density waves can be launched by unseen substellar companions
of $\gtrsim M_{\rm jup}$ \citep{Muto-2012}. \cite{Juhasz-2015} studied
the observability of spirals launched by embedded planets, suggesting
that they are the result of changes in the vertical scale height of
the disk rather than density perturbations \citep[see
  also][]{pohl-2015}. Interestingly, the large pitch angles in spirals
with $m=2$ symmetry can be explained by the presence of massive
planets {\it exterior} to the spiral features \citep{Dong-2015a}. An
origin in gravitational instabilities (GI) is also plausible for massive
disks, but limits the size of the spirals to $\lesssim$100~AU
\citep[e.g.][]{Lodato-Rice-2004, Dong-2015}. Thus current spiral
models require either massive planets, or gravitationally unstable
disks, or both \citep{pohl-2015}.

Motivated by the recent identification of deep shadows cast by an
inner warp in HD~142527 \citep{Marino-et-al-2015,
  Casassus2015ApJ...811...92C}, in this letter we consider the
dynamical consequences of the temperature forcing on the outer disk as
it periodically flows under such shadows.  We use 2D hydro simulations
to follow the evolution of a passive gaseous disk subjected to
non-axially symmetric shadowing, and report on the development of
spiral waves (Sec.~\ref{model}). Based on these results
(Sec.~\ref{results}), we propose an alternative mechanism to trigger
spirals from illumination effects in the outer regions of gapped
systems (Sec.~\ref{conclusions}).

\section{The Model}\label{model}

We are interested in the evolution of a self-gravitating, planetless,
gaseous circumstellar passive disk. The stellar radiation field is
fixed to two reference values, $L_\star = 1 \rm L_\sun$, and $L_\star
= 100 \rm L_\sun$. We consider disks with masses $M_{\rm d} = 0.05$,
and $M_{\rm d} = 0.25 \rm M_\star$, without including viscous
dissipation, which translates into very low accretion rates. A
constant kinematic viscosity prescription is used, given by $\nu = 4.5
\times 10^{7} \rm m^2 ~ s^{-1}$. 
The evolution of the disk is followed for about $10^4$ years.

Using three-dimensional radiative transfer calculations of warped  disk structures,   
  \cite{Whitney2013}  shows that, in general,  tilted thick inner disks
  will cast point-symmetric shadows onto the outer disk. Accordingly,
our model features two point-symmetric shadows projected along the
disk with opening angles $\delta \sim 28^\circ$.  This value
  is motivated by HD~142527's case reported by
  \citet{Marino-et-al-2015}. 
  
The simulations were performed with the public two-dimensional
hydrodynamic code {\sc
  fargo-adsg}\footnote{http://fargo.in2p3.fr/spip.php?rubrique9}
\citep{Baruteau-Masset-2008}, after implementing a non-stationary
energy equation that includes a blackbody radiative cooling. The code
solves the Navier-Stokes, continuity and energy equations on a
staggered mesh in polar coordinates $(r,\phi)$. For detailed description
of the code see \cite{Masset-2000} and \cite{Baruteau-Masset-2008}.

\subsection{Code units and initial setup} \label{units}

The simulations were run in code units, assuming
the central star mass, $M_0 = M_{\star}$,
and, $r_0 = 1$~AU, as the units of mass and
length. The code unit of time, $t_0$, corresponds to the orbital
period at $r_0$ divided by $2\pi$, that is $t_0=(G M_{\star} /
r_{0}^3)^{-1/2}$. The gravitational constant is $G=1$ in code units. 
The temperature unit  is $G M_{\star} \mu m_p / (k_B r_{0})$, with $\mu$ being the mean
molecular weight ($\mu = 2.35$ in all our simulations), $m_p$ the
proton mass and $k_B$ the Boltzmann constant. We adopt two values for
the disk-to-star mass ratio, $q = 0.25$ and $0.05$.

The computational domain in physical units extends from $r = 10$ to
$150$~AU over $n_r = 400$ logarithmically spaced radial cells. The
grid samples $2\pi$ in azimuth with $n_\phi = 800$ equally spaced
sectors. Gas material is allowed to outflow at the disk edges.

The initial density profile scales with $r^{-1}$:
\begin{equation}
\Sigma(r) = \Sigma_0 \left(\frac{r_0}{r}\right),
\label{sigmaScale}
\end{equation}

\noindent where the value of $\Sigma_0$ is set by fixing the disk mass
to 0.05 and 0.25 $\rm M_\star$ ($M_{\rm disk} = \int_{r_{\rm
    min}}^{r_{\rm max}} 2 \pi \Sigma(r) r dr$). The disk initial
aspect ratio, $H/r$, is fixed to 0.05.  All models include
self-gravity. Our choice of parameters is
similar to other numerical models of transition disks
\citep[e.g.][]{Dipierro-2015}.

\subsection{Implementation of shadows}\label{shadow_modeling}

The stellar irradiation heating per unit area is given by
\citep{Frohlich-2003}:

\begin{equation}
Q^+_\star = (1 - \beta) \frac{L_\star}{4 \pi r^2} \cos\phi,
\label{Qstar}
\end{equation}

\noindent where $\beta = 0.1$ is a reflection factor (albedo) 
and $\phi$ is the angle formed between the
incident radiation and the normal to the surface, given by $\cos \phi
\simeq \frac{dH}{dr} - \frac{H}{r}$. The disk scale height is assumed
to be in hydrostatic equilibrium, i.e., $H = r~c_s(T)/v_k$, with
$c_s(T)$ the sound speed and $v_k$ the Keplerian velocity. $L_\star$
is the stellar luminosity.

%%%%%%%%%%%%%%%%%%%%%%%%%%%%%%%%%%%%%%%%
\begin{figure}[t]
%\epsscale{1.2}
\plotone{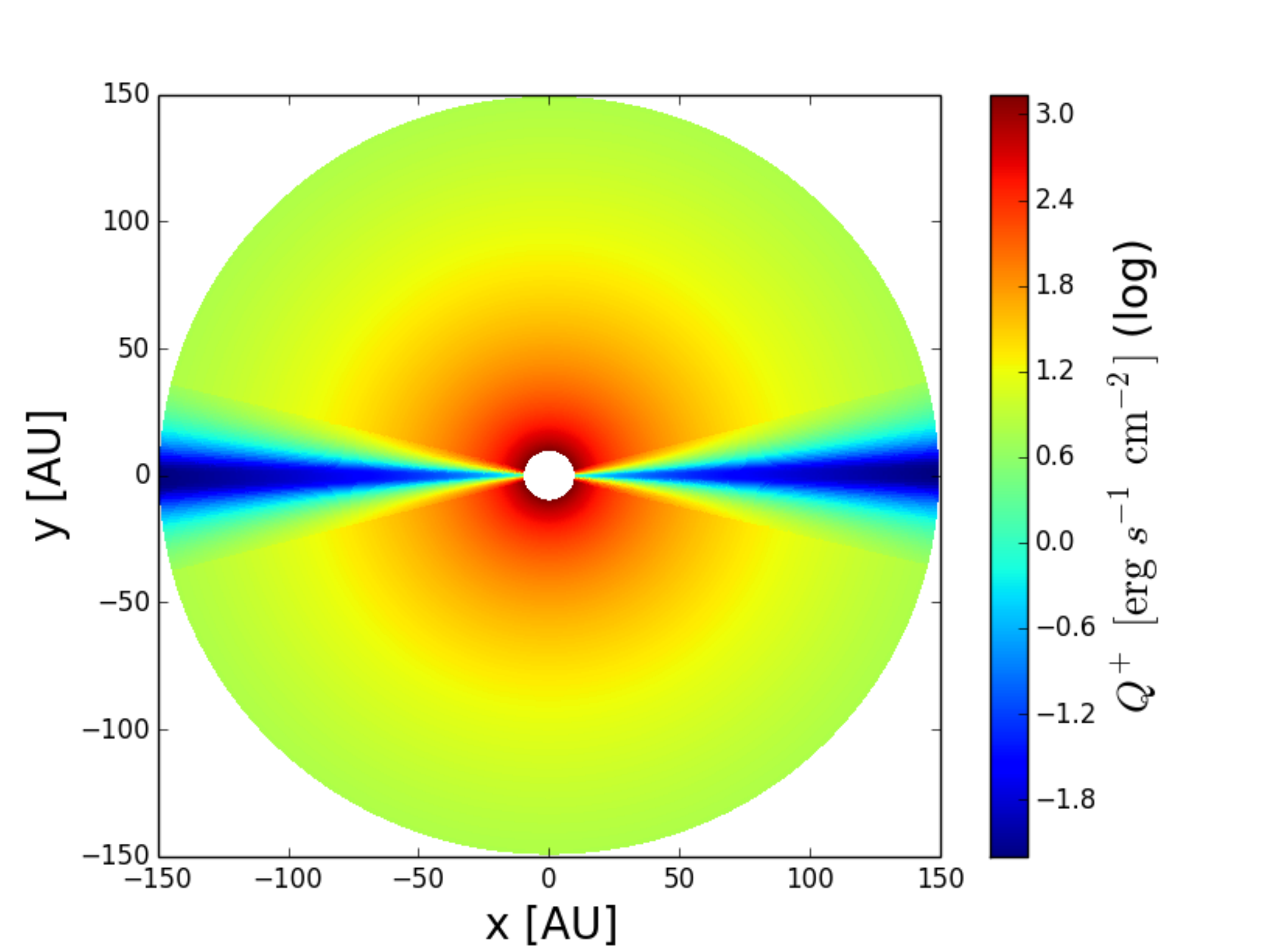}
\caption{Initial profile for the stellar heating rate per unit area
  $Q^+_d(r, \phi)$ for a model with $L_\star = 1 L_\sun$. The shadows
  subtend $0.5~{\rm rad}$ or 28.6$^\circ$. The plot is in log scale
  with units  $\rm erg ~ s^{-1} ~ cm^{-2}$.}
\label{Qplus}
\end{figure}
%%%%%%%%%%%%%%%%%%%%%%%%%%%%%%%%%%%%%%%% 

In our simulations, shadows are cast by an inner region (inside the
computational domain) which blocks a fraction of the stellar
irradiation $Q^+_\star$ within an angle $\delta$. The irradiation
heating per unit area, including shadows projected onto the disk,
$Q_d^+(r, \phi)$, reads;

\begin{equation}
Q^+_d(r, \phi) = \left\{ \begin{array}{ll}
    F(t) f(\phi) Q^+_\star &\mbox{ if $|\phi| > (\pi - \delta/2) \land
  |\phi | < \delta/2$} \\
 Q^+_\star    &\mbox{ otherwise },
       \end{array} \right.
\label{QstarFinal}
\end{equation}
where illuminated regions are connected by a smooth azimuthal
function $f(\phi) = 1 - A \exp{(-\phi^4/\sigma^2)} - A \exp{(-(\phi -
  \pi)^4/\sigma^2)}$, with $A=0.999$ and
$\sigma=\delta/10$. The
time-dependent function $F(t)$ is a shadow-tapering factor which
gradually enables the shadows over a timescale of 30 orbits\footnote{from the used code units, one orbit correspond to one year.}.
Figure~\ref{Qplus} shows the stellar irradiation prescription
described by Equation~\ref{QstarFinal}.

\subsection{The energy equation}

The equation for the thermal energy density, $e$, reads (e.g.,
\cite{D'Angelo-2003}):
\begin{equation}
\frac{\partial e}{\partial t} + \overrightarrow{\nabla} \cdot (e
\overrightarrow{v}) = -P \overrightarrow{\nabla} \cdot
\overrightarrow{v} + Q^+_d - Q^-,
\label{energy1}
\end{equation}
where $\overrightarrow{v}$ is the gas velocity, $P$ the pressure, and
$Q^+_d$ the stellar heating rate per unit area described by
Eq.~\ref{QstarFinal}. We implemented a radiative cooling per unit area
function, $Q^- = 2 \sigma T^4 / \tau$, where $\sigma$ is the
Stefan-Boltzman constant and $\tau$ the optical depth given by $\tau =
\case{1}{2} \Sigma \kappa$. A constant opacity given by an effective
$\kappa = 130 \rm ~cm^2 ~ g^{-1}$, obtained by taking into account an average opacity for a mix of dust species,
informed by spectral energy distribution fitting of HD 142527 \citep{Casassus-et-al-2015}. Dependency on opacity is further discussed in Section \ref{results}. 
For more details about this energy prescription, see \cite{2015ApJ...806..253Montesinos}.

To close the system of equations, an ideal equation of state is used,
\begin{equation}
P = \Sigma T \overline{R},
\label{state1}
\end{equation} 
where $T$ is the mid-plane gas temperature and $\overline{R} = k_B /
\mu m_p $ is the gas constant. The thermal energy density is related
to the temperature through
\begin{equation}
e = \Sigma T \left(  \frac{\overline{R}}{ \gamma - 1}  \right),
\label{state2}
\end{equation}
where the adiabatic index is fixed to the diatomic gas value $\gamma =
1.4$.

\begin{figure*}[]
\centering
\includegraphics[scale=0.25]{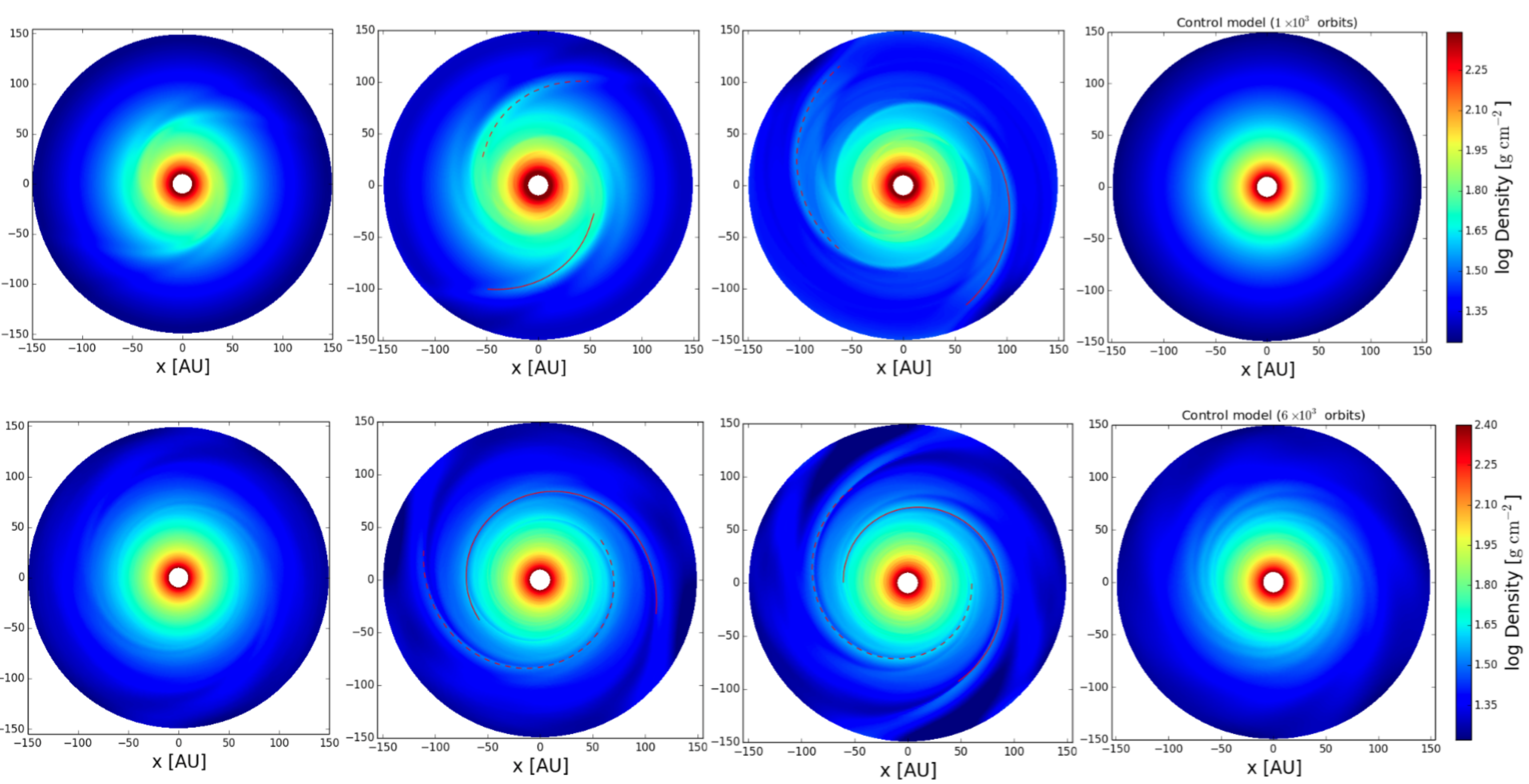}
\caption{Density field evolution of a model disk with $M_{\rm d} =
  0.25 \rm M_\star$.  \textbf{Top row}: $L_\star = 100 \rm L_\sun$
  model. From left to right; 150, 250, 500 orbits, respectively.
  \textbf{Bottom row}: $L_\star = 1 \rm L_\sun$ model (same disk
  mass). From left to right; 2500, 3500, and 4000 orbits.  The upper
  rightmost panel is a control simulation (i.e., an unshadowed model)
  with $L_\star = 100 \rm L_\sun$ after 1000 orbital periods.  In this
  case, no azimuthal structures appear during the disk's
  evolution. The bottom rightmost panel corresponds to a control run
  with $L_\star = 1\rm L_\sun$ in which the first structures appear
  caused by gravitational instabilities after 6000~$t_0$. }
\label{Dens1}
\end{figure*}

\section{Results}\label{results}

\subsection{Spiral structures in the density field}\label{spiral-structure}

Control simulations without shadows were performed in order to test
for azimuthal structures unrelated to illumination effects.
Additionally, random noise at the $\sim 0.1\%$ level was injected to
the initial surface density of these simulations in order to test for
stability and fragmentation.

Azimuthal features due solely to the disk self-gravity develop in
control runs with $M_{\rm d}=0.25 \rm M_\star$ and $L_\star= 1 \rm
L_\sun$, on timescales of $\gtrsim 10^3$ orbits. This is expected for
such gravitationally unstable disks \citep{Dipierro-2015}. For control
models with $M_{\rm d}=0.05 \rm M_\star$ (gravitationally stable), no
spiral structures emerge, independent of the strength of the stellar
irradiation field.

When shadows are enabled, spiral-like structures emerge in the surface
density of a simulated 0.05~M$_\star$ disk irradiated by a $100 \rm
L_\sun$ star. The onset of these structures occurs approximately after
130 orbital periods. On the other hand, no spirals appear if the
0.05~$M_\star$ disk is illuminated by only $1 \rm L_\sun$.

We consider the onset of the shadow-induced spirals to be the time at
which their scale height suffices $\delta H/H \simeq 0.3$. This
criterion is evaluated within the first 10-20~AU radius from the star. The
perturbation on the scale height is calculated as $\delta H \equiv H -
H_0$, where $H_0$ represents the background scale height without
shadows. We note that a relative change of $\sim 0.2$ in scale height
is sufficient to produce detectable azimuthal signatures in scattered
light predictions \citep{Juhasz-2015, Dong-2015a}.

For the more massive 0.25~M$_\star$ disk, with $L_\star = 1 \rm
L_\sun$, the first non-axisymetric structures appear approximately
after 2500~orbits. For the same disk mass, but increasing stellar
irradiation to 100~L$_\sun$, the first spirals arise shortly after
only 150~orbits. These different timescales will be discussed in the
next subsection.

Figure~\ref{Dens1} shows the impact of different stellar irradiation
values on the evolution of the density field, for a 0.25~M$_\star$
disk.  Top row shows the 100~L$_\sun$ model after 150, 250, and
400~orbits (from left to right, respectively), while bottom panels
show the 1~L$_\sun$ model at 2500, 3500, and 4000~orbits (from left to
right). Each first snapshot corresponds to the moment spirals appear
for the first time, as defined above.  The rightmost top and bottom
panels correspond to control runs, in which no shadows are present.

In order to test the dependency of 
the disk evolution with the opacity, we also explore the case with $\kappa = 1 \rm ~cm^2 ~ g^{-1}$, finding the same results.
This insensibility to the opacity, in order to produce shadow-induced spiral arms, is not surprising. The radiative cooling 
is computed according $T^4/\tau$, therefore, changes in $\kappa$ implies (roughly speaking) changes in the temperature field of about $\sim \kappa^{1/4}$. Moreover,
the amplitude of a thermal perturbation responsible for spiral development is given by  $\delta T / T$ (see next section), 
therefore, in an equilibrium situation in which $Q^+ \sim Q^-$, the assumption of a constant opacity results in a thermal amplitude perturbation  independent of
the opacity choice. 

Our 2D simulations are not radiation hydrodynamics but just an ideal approximation.  
The role of the opacity will be considered in a future work,  where we will study 
the illumination-induced spirals using full 3D radiation hydrodynamic simulations.

\subsubsection{Pitch angle}

Spiral arm's pitch angles were computed by fitting the spirals with an
Archimedean equation $r(\phi) = A_0 + A_1 \phi^n$. The pitch angle
$\Pi$ is then obtained through $\tan{\Pi} = (1/r) dr/d\phi$. From this
parameterized curve, a \textit{global pitch angle} is calculated as
the mean value along the curve.

$\Pi$ varies over time for different model parameters. The
shadow-induced spiral arms extend from $A_0$ ($\sim$60-90~AU, see
Table\ref{spiral_table}) to the outer region of the disk $\sim$150~AU
(see Figure \ref{Dens1}), as opposed to spirals resulting from
gravitational instabilities which persist over scales $\lesssim
100$~AU \citep[e.g.,][]{Dong-2015a}. Table~\ref{spiral_table}
summarizes the most important features of the shadow-induced spirals.

\begin{deluxetable*}{lccccc}[t]
\tabletypesize{\scriptsize}
%\rotate
\tablecaption{Main spiral and disk characteristics}
\tablewidth{0pt}
\tablehead{
\colhead{Model parameters}  &  \colhead{Orbit number} & \colhead{Spiral parameters} & \colhead{$\rm Toomre^a$} & \colhead{Birth time-scale of  }\\
\colhead{ }  &  \colhead{ } & \colhead{$A_0$, $A_1$, $n$, $\Pi^\circ$} &  \colhead{$Q_{\rm min}$} & \colhead{the $\rm structures^b$ (in orbits)   }
}
\startdata
\vspace{-0.2cm} $M_d=0.05 \rm M_\star$; $L_\star = 1 \rm L_\sun$ & 3500  &        n/a                   &  1.1    & n/a  \\
\\
\vspace{-0.2cm}                                                                                  &  4000  &        n/a                   &  1.1  & \\
\\ \hline \\[-1.8ex]
\vspace{-0.2cm} $M_d=0.05 \rm M_\star$; $L_\star = 100 \rm L_\sun$   & 250  &  68.7; 14.6; 2.1; $22.59^\circ \pm 3.5$   &  3.4 & $\sim 150$    \\
\\
\vspace{-0.2cm}                                                                                  &      500  & 56.9; 4.2; 1.64;  $13.82^\circ \pm 3.5$        &  2.6  & \\
\\\hline \\[-1.8ex]
\vspace{-0.2cm} $M_d=0.25 \rm M_\star$; $L_\star = 1 \rm L_\sun$  & 3500  & 70.3;  4.6;  1.68; $13.82^\circ \pm 3.5$  &  0.6   & $\sim 2500$  \\
\\
\vspace{-0.2cm}                                                                                   &  4000  & 60.3; 5.4;  1.45; $11.35^\circ \pm 3.5$ &  0.5 &  \\
\\\hline \\[-1.8ex]
\vspace{-0.2cm} $M_d=0.25 \rm M_\star$; $L_\star = 100 \rm L_\sun$   & 250  & 60.3;  28.2; 1.35; $15.04^\circ \pm 2.5$ &  1.0 & $\sim 150$  \\
\\
\vspace{-0.2cm}                                                                                       & 500  & 87.2;  17.8; 1.43; $13.09^\circ \pm 2.0$ &  1.2 &   \\

\enddata
%% Text for table notes should follow after the \enddata but before
%% the \end{deluxetable}. Make sure there is at least one \tablenotemark
%% in the table for each \tablenotetext.

\tablecomments{The spirals can be described with the Archimedean
  function $ r(\phi) = A_0 + A_1 \phi^n$, where the pitch angle is
  given by $\tan \Pi = (1/r) dr/d\phi$.  The value $A_0$ (in AU) gives
  the inner radius of the spirals, which in our models extend to the
  outskirt of the disk ($\sim 150$ AU).}

\tablenotetext{a}{Minimum local value of the Toomre parameter at the
  given Orbit number.}

\tablenotetext{b}{Orbit number when spirals are distinguishable for
  the first time in the density field (see subsection
  \ref{spiral-structure}).}

\label{spiral_table}
\end{deluxetable*}

\begin{figure*}[t]
%\epsscale{1.5}
\centering
\includegraphics[trim=2.5cm 0.0cm 1.0cm 0.0cm, clip=true,width=0.3\textwidth]{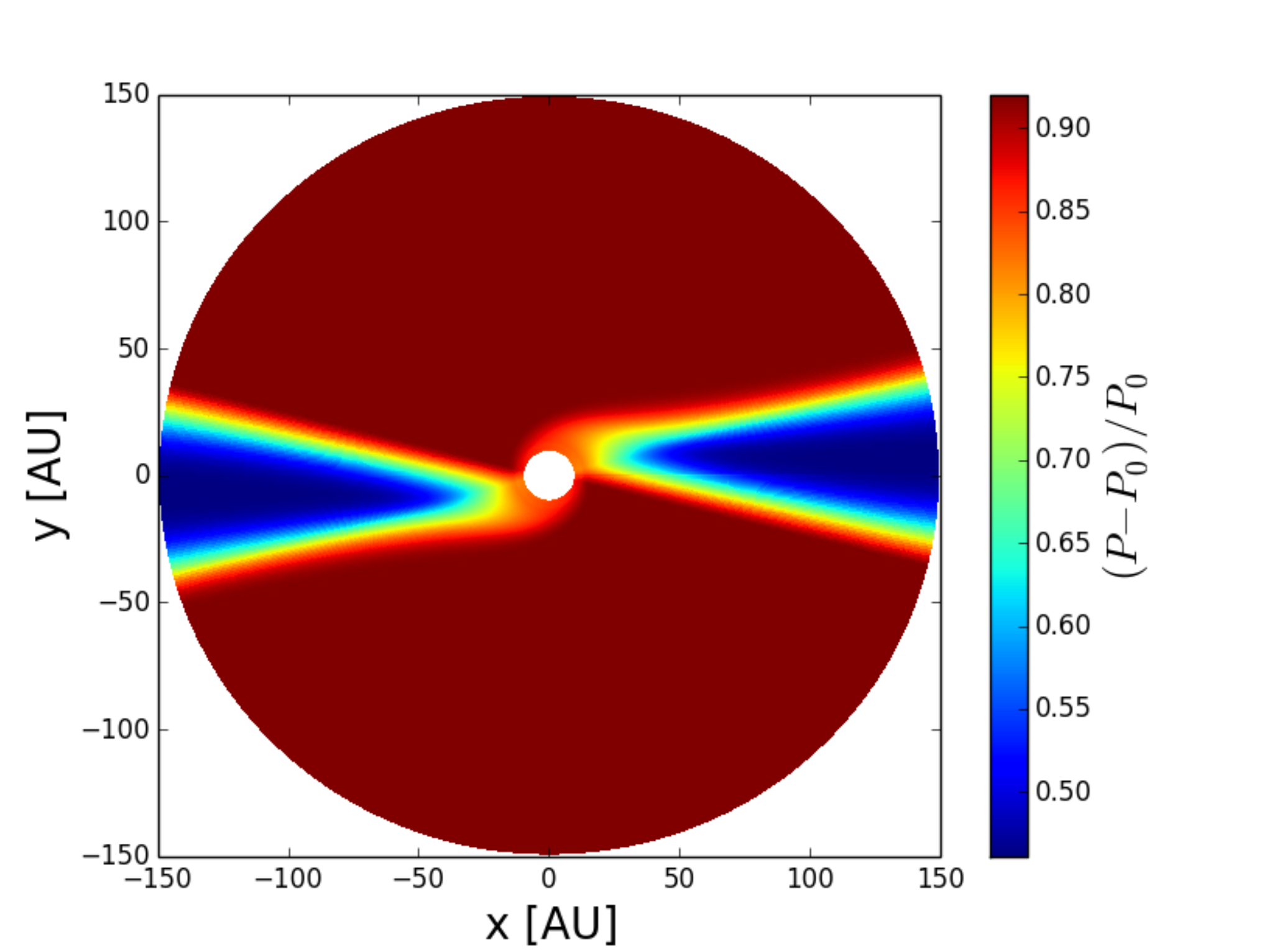} 
\includegraphics[trim=2.5cm 0.0cm 1.0cm 0.0cm, clip=true,width=0.3\textwidth]{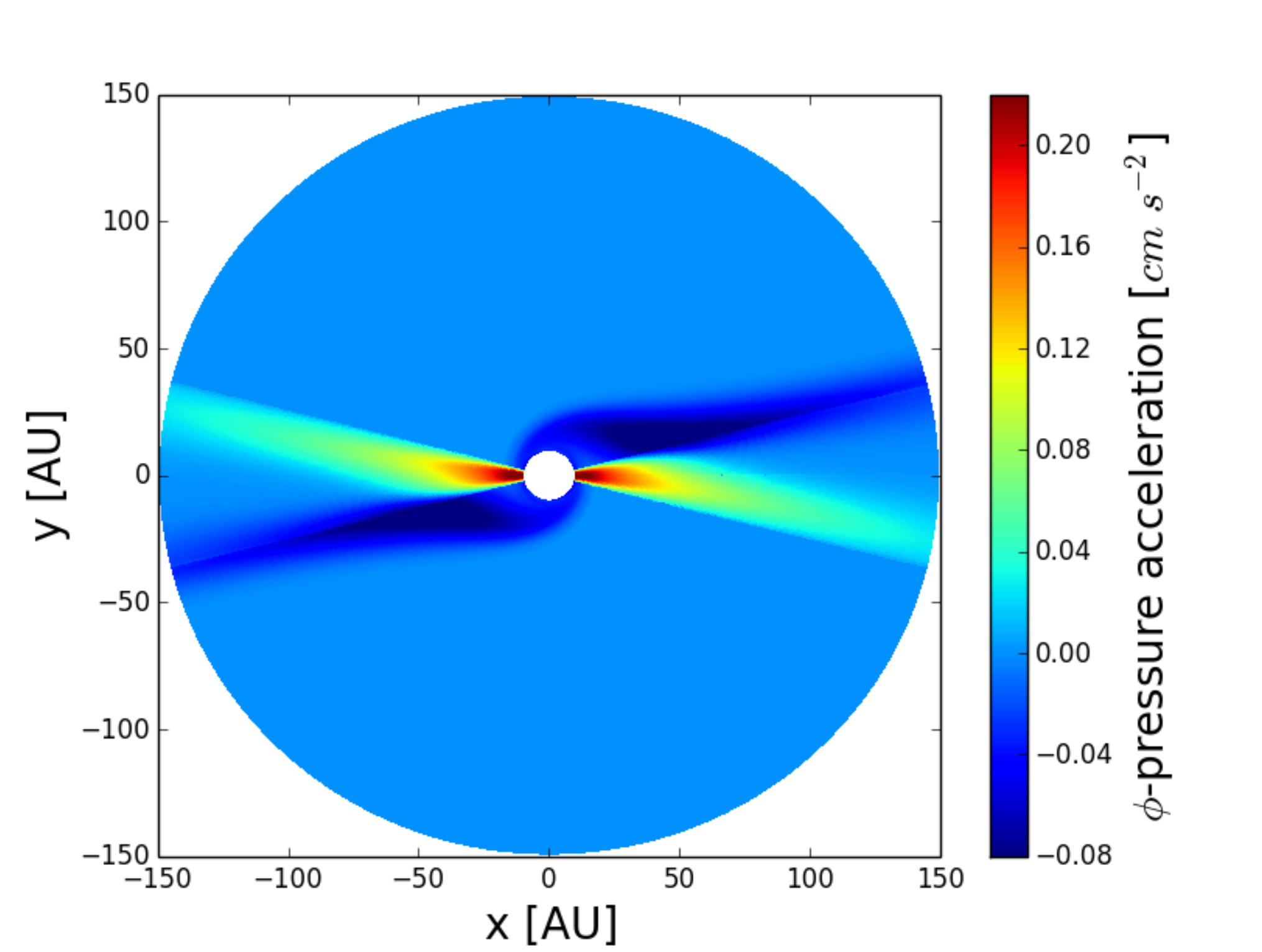}
\includegraphics[trim=2.5cm 0.0cm 1.0cm 0.0cm, clip=true,width=0.3\textwidth]{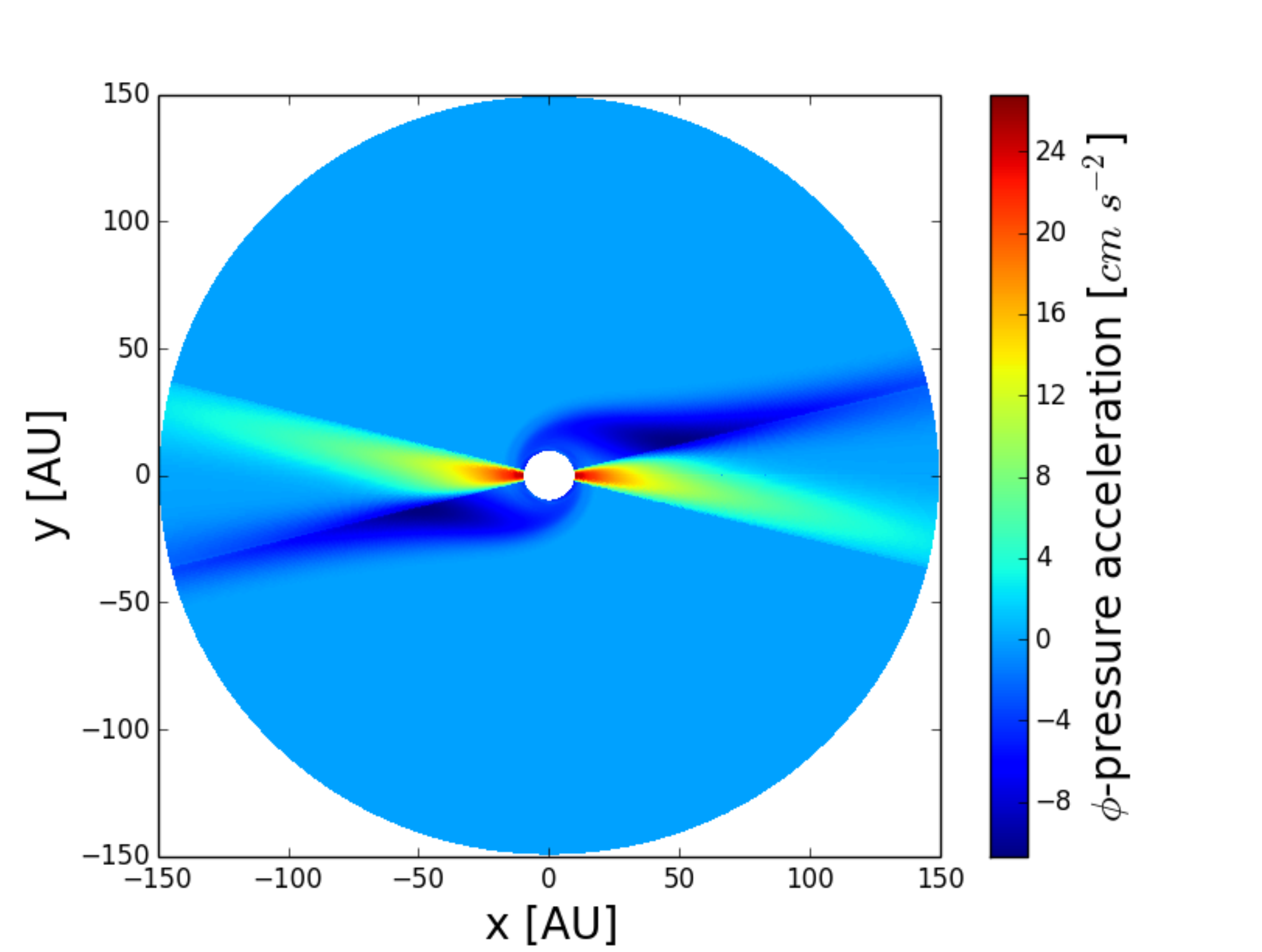}

\caption{Left panel shows how the illumination effect impacts the
  pressure field (i.e., $(P - P_0) / P_0$, where $P_0$ is the
  background pressure field without the shadows) after 30 orbits for a
  model disk with $L_\star = 100 \rm L_\sun$ and $0.25 \rm M_\star$.
  Pressure decreases at the shadows' positions. Middle and right
  panels show azimuthal pressure acceleration ($a_\phi = -(1/ r\Sigma)
  \partial P/\partial \phi $) after 30 orbits for models with $L_\star
  = 1 \rm L_\sun$ and $L_\star = 100 \rm L_\sun$, respectively (same
  disk mass for both models).}
\label{grad}
\end{figure*}

\subsection{Formation of spiral-like structures from shadows} 

Perturbations in a differentially rotating gaseous disk tend to wind
up into spiral patterns (e.g., see the control model, bottom-rightmost
panel in Figure~\ref{Dens1}). In our models, projected shadows act as
forcing perturbations, which efficiently and independently of the
presence of random source of symmetry breaking (e.g., gravitational
instabilities), produce spirals. Because of the point-symmetric nature
of this perturbation, density waves are excited with $m=2$ morphology.
Each shadow appears to trigger one spiral arm.

Shortly after the shadows have been cast onto the disk, and before the
disk thermalizes, the gas pressure (Eq.~\ref{state1}) plummets at the
shadowed regions.  Figure~\ref{grad} (left panel) shows the relative
change in pressure $(P - P_0) / P_0$, where $P_0$ is the background
unshadowed gas pressure. This snapshot corresponds to a 100~L$_\sun$
and $0.25 \rm M_\star$ model disk immediately after shadows are
enabled. Figure~\ref{grad} shows the azimuthal pressure acceleration,
$\vec{a}_\phi$, after 30 orbits for two $0.25 \rm M_\star$ disk models
with $L_\star = 1\rm L_\sun$ and $100 \rm L_\sun$, middle and right
panels, respectively. The pressure acceleration magnitude increases
with stellar irradiation: the maximum value for the pressure
acceleration increases from $0.2$ to $24 ~ \rm cm ~ s^{-2}$ when the
stellar irradiation passes from $L_\star = 1 \rm L_\sun$ (middle) to
$L_\star = 100 \rm L_\sun$ (right panel).

As the gas in the disk flows around the star (in anticlockwise
direction for all figures), it periodically enters and exits shadowed
regions. At the interfaces between shadowed and fully illuminated
sections, the gas is subjected to an \textit{azimuthal acceleration}
due to pressure gradients (i.e., $\vec{a}_\phi = -(1/r \Sigma(r))
\nabla_\phi P $). As the gas enters a shadowed region (coming from
high to low pressure), it experiences a positive azimuthal
acceleration (anticlockwise). On the other hand, as it exits the
shadow (moving from low to high pressure), the gas experiences a
negative azimuthal acceleration (clockwise). The difference between
the gain and loss of azimuthal acceleration is uneven, resulting in a
net azimuthal acceleration in the anticlockwise direction (gas
rotation direction), which varies with radius within the first
$\sim$50~AU.  This creates a strong source of axisymmetry breaking,
locally pushing the gas to move faster around the central star at the
inner regions of the disk, piling up material at the shadows frontier
close to this inner region, which then leads to the
formation of a trailing mode spiral-like structures outspread by differential
rotation. 
 
Gas pressure acceleration depends on stellar luminosity as more
luminous stars feed more energy into their disks, rising the
temperature of the illuminated sections while leaving the shadowed
region unaffected. This produces a higher contrasts in the pressure
field. This is the reason why spirals appear early ($\sim 130$ orbits)
in models with $L_\star = 100 \rm L_\sun$, when compared to 1~L$_\sun$
models in which the first spirals emerge only after $\sim 2500$
orbits. The key factor is the contrast in pressure between dark and
illuminated regions.

Depending on the gas cooling time, thermal perturbations can wear off over time. 
As the disk thermalises, the illumination pattern becomes less effective at forcing 
a perturbation on the gas. In our simulations, this tendency to homogenize is seen after $10^5$ orbits. 
Ultimately, the induced spiral arms tend to blend and lose contrast with the background gas unless 
gravitational instabilities kick in. In these large viscous time-scales, GI induced spiral structures are expected to remain quasi-steadily (e.g. Dipierro et al. 2015).

\subsection{Gravitational instabilities}\label{GI}

Gravitational instabilities can be locally characterized by the Toomre
parameter\footnote{For \textit{extremely} massive disks, $\Omega$
  should be replaced by the \textit{epicyclic frequency $\kappa$},
  defined by $\kappa^2 = (1/r^3)d/dr(r^4\Omega^2)$. In our case,
  $\kappa \approx \Omega$.} given by $Q = \frac{c_s \Omega}{\pi G
  \Sigma}$, where $c_s$, $\Omega$, and $\Sigma$ correspond to sound
speed, angular velocity and surface density, respectively
\citep{Toomre-1964}. The disk becomes unstable when $Q < Q_{\rm
  crit}$, where $Q_{\rm crit}$ defines the range for which a disk is
marginally unstable: $ 1 \lesssim Q_{\rm crit} \lesssim 2$.
 
According to the local Toomre parameter, denser and colder regions of
the disk tend to have lower $Q$ values. The $M_{\rm d} = 0.05 \rm
M_\star$ and $L_\star = 100 \rm L_\sun$ model is stable everywhere in
the disk, with a minimum $Q$ value of 3.35. The most \textit{unstable}
model ($M_{\rm d} = 0.25\rm M_\star$ and $L_\star = 1 \rm L_\sun$
attains a minimum local value $Q=0.5$. It is important to remark that
spiral structures emerge even if we disable self-gravity from our
simulations. The shadow-induced spirals do not require the presence of
gravitational instabilities for their development.

\subsection{Radiative transfer}

We input our simulations into a 3D radiative transfer code to produce
scattered light predictions in $H$-band ($1.6 \mu m$). We used the
{\sc RADMC3D}\footnote{http://www.ita.uni-heidelberg.de/$\rm
  \sim$dullemond/software/radmc-3d/} code (version 0.39), assuming
that the dust follows the same density field as the gas in the
simulation, with a gas-to-dust ratio of 100. 
The dust distribution model consists of a mix of two
common species: amorphous carbon and astronomical silicates. We used
Mie model (homogeneous spheres)  to
compute dust opacities for anisotropic scattering (Bohren, C. \& Huffman, D. R. 1983). 
The optical constants for amorphous carbon (intrinsic grain densities
2~g~cm$^{-3}$) were taken from \cite{Li-Greenberg-1997}, and for
silicates (intrinsic grain densities 4~g~cm$^{-3}$) from
\cite{Draine-Lee-1984}.

To produce a 3D volume, the vertical density structure was solved
assuming hydrostatic equilibrium. The disk scale height is obtained
from the temperature field (Section \ref{shadow_modeling}).

The temperature in the vertical direction is assumed to be constant and
equal to the midplane temperature with $T_{\rm gas} = T_{\rm dust}$. 
The final image prediction is
rendered using a second order volume ray-tracing.

\begin{figure}[t]
%\epsscale{1.0}
\plotone{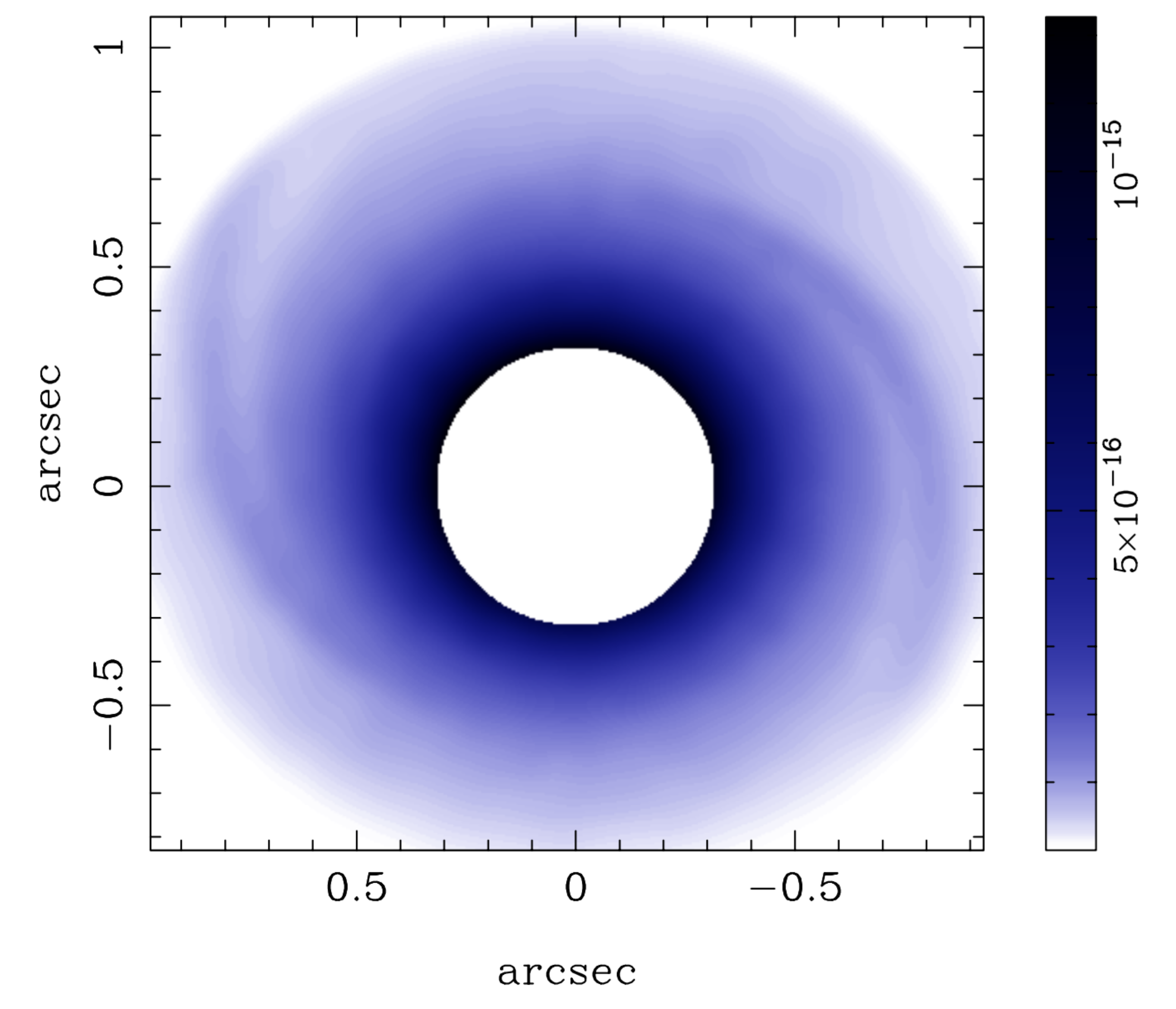} % band_H_spiral.eps}
\caption{Scattered light prediction for $H$ band ($1.6 ~ \mu m$)
  obtained from 3D radiative transfer calculations. The hydrodynamical
  input corresponds to a model with a disk mass of 0.25 $ \rm
  M_\star$, and $1 \rm L_\sun$ star. Color stretch
  is linear, with units $\rm erg ~ cm^{-2} s^{-1} Hz^{-1}
  ster^{-1}$.}
\label{radmc}
\end{figure}

Figure \ref{radmc} shows a model prediction in H band
based on a 0.25~M$_\star$
disk with stellar irradiation of 1~L$_\sun$, after it had evolved for
3500~orbits. We recover the spiral arms in scattered light. We also
obtain observable spiral structures in $L$-band. The scattered light
spirals follow the same equation as the one obtained from the density
field, i.e., $\Pi = 13.82^\circ$ (see Figure \ref{Dens1}, bottom row,
orbit 3500, and/or Table \ref{spiral_table}).

 It is worth mentioning that the assumption of  \textit{hydrostatic equilibrium} in  
 two-dimensional models likely underestimates the contrast of spirals images
when compared with with full 3D hydrodynamical models (Zhu et al. 2015). In that 
case, the illumination effects impacting the disk should produce even more prominent scattered light images
of spirals than those reported here.

%%%%%%%%%%%%%%%%%%%%%%%%%%%%%%%%%%%%%%%%
%%%%%%%%%%%%%%%%%%%%%%%%%%%%%%%%%%%%%%%%

\section{Concluding remarks}\label{conclusions}

We investigated the hydrodynamical consequences of illumination
effects in transitional disks, such as shadows cast onto a
circumstellar disk, focusing on the development of spiral
structures. The evolution of a passive disk was calculated using 2D
hydro-simulations including stellar irradiation values of 1 and 100
$\rm L_\sun$, and with two point-symmetric shadows cast within an
opening angle of 28$^\circ$ for disk masses of $M_{\rm d} = 0.05$ and
$M_{\rm d} = 0.25$~M$_\star$.  Pressure gradients due to temperature
differences between obscured and illuminated regions induce spiral
structures in the density field.  These spirals emerge independent of
whether self-gravity is enabled or not. The structures observed in the
density field can be characterized by an Archimedean spiral which
attain nearly constant pitch angles $\sim$11$^\circ$-14$^\circ$, and
extend from $\sim$60~AU to the outer disk rim. Radiative transfer
calculations, for a model with $0.25 \rm M_\star$ disk mass and
stellar irradiation of $1 \rm L_\sun$, predict that these
shadow-induced spirals should be detectable in $H$-band scattered
light images.

Recently, \cite{Rafikov2016} presented an analytical study of the
effect of density waves on proto-planetary disks.  They found that
density waves with contrast of order unity, such as the ones resulting
from our models, produce an enhancement in the accretion rate that
could explain the large cavities observed in some transition disks.
In HD~142527, the inner disk casts a shadow that explains the observed
dips in the outer disk \citep{Marino-et-al-2015}, and, as shown in
this letter, drives spiral waves \citep{Casassus2012ApJ...754L..31C,
  Christiaens-2014}. Following \cite{Rafikov2016}, the enhanced
accretion produced by these spirals could have depleted the inner
$\sim$100~AU of this disk. The HD~100453 disk
\citep[][]{Wagner2015ApJ...813L...2W}, which we suggest bears strong
similarities to HD~142527 as spirals are seen to stem away from dark
regions in the outer disk, might also fit in this scenario.  A more
detailed study of the interplay between the different processes will
be investigated in a follow-up paper.

\section*{Acknowledgments}

We thank Cl\'ement Baruteau for very useful comments on this
paper. Financial support was provided by Millennium Nucleus grant
RC130007 (Chilean Ministry of Economy). M.M. acknowledges support from
CONICYT-Gemini grant 32130007. S.C, S.P. and J.C. acknowledge
financial support provided by FONDECYT grants 1130949, 3140601 and
1141175. The authors also thank the referee for her/his suggestions 
that have improved this letter.

%\bibliography{astro}

\end{document}